# A GENERAL XML-BASED DISTRIBUTED SOFTWARE ARCHITECTURE FOR ACCESSING AND SHARING RESOURCES

Samuel Cruz-Lara, Patrice Bonhomme, Christophe de Saint-Rat and Laurent Romary
*LORIA (UMR 7503) Laboratoire Lorrain de Recherche en Informatique et ses Applications*
*CNRS – INRIA – Universities of Nancy*
*FRANCE*
*{Samuel.Cruz-Lara, Patrice.Bonhomme,  Laurent.Romary, desaintm}@loria.fr*

## Abstract

*This paper presents a general xml-based distributed software architecture with the aim of accessing and sharing resources in an opened client/server environment. The paper is organized as follows: First, we introduce the idea of a "General Distributed Software Architecture". Second, we describe the general framework in which this architecture is used. Third, we describe the process of information exchange and we introduce some technical issues involved in the implementation of the proposed architecture. Finally, we present some projects which are currently using the proposed architecture in the domain of textual resources.*

## 1. Introduction.

Our main objective is to define a general distributed software architecture (see Figure 1) through which it might be possible to access and to share resources, which would be spread among different servers.

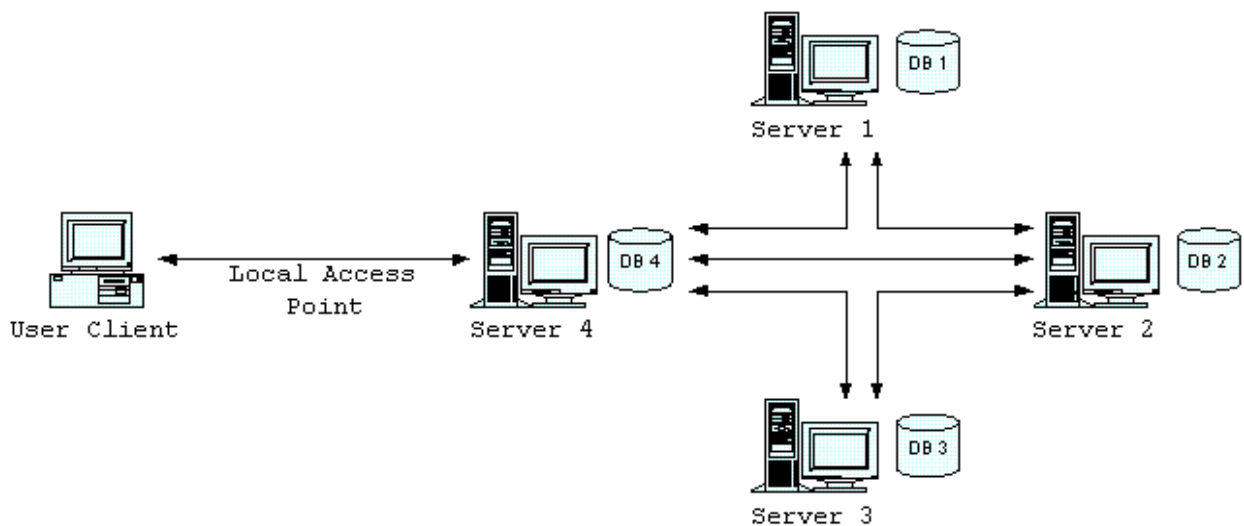

**Figure 1. "General Architecture of the Network".**

There are different reasons why to go about in this way. First, it should be considered that there already exist several sites all around which actually act as brokers for different types of resources which are stocked in some kind of databases. The general idea is that one should be able to bring together all the existing resources, in order to provide any user with a global access to these. In some cases this will lead to more coherence between these databases since redundancies might be detected or complementary resources put into correspondence.

Second, the classical view of a centralised database containing all the information in a given domain is far from applicable to most kind of resources, because of their intrinsic diversity, for example. Also, there is a need for these resources to be created, and above all, maintained at a place where there is the competence to do so. For instance, textual resources may correspond to a wide variety of languages, genres and temporal periods which which are more easily maintained by groups of scholars possessing the proper competence.

Finally, there can be specific constraints that can preclude some given resources to be deported to another site than the site which has originally created them. For example, we can have strong conditions related to property rights on the actual distribution of electronic versions of some textual resources. It is thus more sensible not to take the risk of hampering the agreement by overly spreading the corresponding contents. In the proposed network, each resource is only accessible through specific queries which can thus be controlled as to their actual applicability.

Beyond that, we have also tried to base our proposals and/or implementations on existing (and even emerging) technologies (i.e. Java, CORBA, XML), so that to ensure some kind of durability to our work.

## 2. A General Framework

From the user's point of view, there should not be much change in the way the various resources funds are to be accessed, which means that whether there are one or several servers should be most transparent to him.

As we will see, adopting a distributed framework, as opposed to the classical view of a centralised database, induces several specific problems for which this paper is trying to provide some plausible answers. Among those, we will have to deal specifically with the problem of broadcasting queries to different servers and conversely combining the corresponding result sets. As an example, statistics can only be dealt with in our distributed architecture if part of the computation is kept on the side of the remote servers and part is carried out locally (on the access server).

All tasks dealing with the user interface should be concentrated on the client side, while the servers should accomplish searching and other computationally intensive operations.

The network has the following characteristics:
1. each server is an autonomous unit containing its own data;
2. each server act as a "broker" and transmit, if needed, the request to other servers in the network which are known to it;
3. one given server is accessible to registered users through a general purpose Java-compatible web browser.

### 2.1. A General scenario

Any authorised user will be provided with an environment which will lead him along the following steps:
1. connection to a local server;
2. user identification;
3. choice of working servers. Given the list of available servers - accessible through the local server - in the network, together with their respective server profiles, the user will select those servers which may provide the proper resources or the proper services (tools) he wants to access or to use ;
4. selection of a subset of resources. Through an iterative process of requests to the selected servers, the user will build up a virtual subset (i.e. by way of pointers to individual resources) upon which he will actually work.

### 2.2. User's side

A user may interact with the network via a simple graphical user interface (in fact, only a general purpose Java-compatible web browser is needed).

This interface is designed with a non-technical user in mind. However, technically advanced features should be available in an intuitive way. Although all system components will have their own interface, due to different functions, they must have the same "look and feel" (i.e. surface and behaviour). This reduces the time the user needs to become acquainted with the network and contributes to the aspect of simplicity.

This user interface will be implemented as a client at the level of which little, not to say no, resource processing is to take place (notion of thin client). Basically, the interface will allow a user to make his different queries and will display result sets according to some specific style-sheets associated with these.

#### 2.2.1. Connection

To be in line with the idea of a decentralised network, we have considered that a given user should only have to be registered at one given site and that no central user database should have to be set up.

### 2.2.2. Workspace

In order for a user to select the servers he wants to work with, he must interact with a "working space". The main purpose of the working space is to provide the user with a graphical interface that allows him to have access to the network and work on it.
The working space will offer the following functions:
- Selection of the working servers;
- Selection of working resources;
- Manipulation of the selected resources;
- Definition of the user's preferences.

### 2.2.2.1. Working Servers Selection

Before any query session (resource selection), a user has the possibility to select among a set of online servers, the server(s) he wants to work with. At any moment of the process, the user can edit his list of working servers and modify it.

### 2.2.2.2. Working Resources Selection

The principle is to restrict progressively the choice of resources in such a way that at the end of the selection process the user only keeps the subset he wants to work with. The user has the possibility to query either the whole network or his selected working servers. For that purpose, the user will have access to a friendly query interface for editing, modifying and sending his queries to his local server.
Another more general way for the user to select his working resources is to browse through the whole set of referenced resources and to select the resources he is interested in.
Once a user has selected his working resources, he can keep them by using the "shopping basket" paradigm. As a matter of fact, one user can handle several shopping baskets, which he can save together with his workspace.

### 2.2.2.3. Manipulation of the selected resources.

As the main goal of this architecture is to give a user an access to a large set of resources, it is most important to provide a working space dedicated to the resources he wants to work with. The user may also use a set of tools allowing him to perform some kind of operations, queries for example, on the selected resources.

## 2.3. Server's side

As we have said, each server is an autonomous unit containing its own data. We have also considered that a given user should only have to be registered at one given site. From the point of view of the network, this results in making a distinction between:
- "specialised servers", to manage users and resources, and
- "broker servers", to share all existing resources.
A given server can bear one or the other status, or even both depending on the actual implementation.

### 2.3.1. Users and Resources

Each server should manage not only its own resource database, but also its own users database. The users database contains both, the general user identification information, and all information related to workspaces.

### 2.3.2. Sharing Resources

Upon requests each server may act as a broker, that is, in order to share all the existing resources, it transmits requests to the other servers in the network. Each time a request is broadcast from an access server to a remote server, an identification tag (user id and authorisation level) should be transmitted in order to evaluate the applicability of the request.

## 2.4. A General Distributed Software Architecture

The network architecture is based on three major actors:

- The Users (or Clients) which have been described in the preceding sections.
- The Network Management Unit (NMU).
  The NMU should be considered as the heart of the network: it allows to link all the servers which are connected to network. The NMU also maintains a database of information related to these servers (i.e. names, addresses, profiles, etc.).
- The servers.
  As mentioned before, this is a set of independent but associated specialised servers

# 3. The "Network Management Unit"

The Network Management Unit (NMU) is the only persistent link between the different servers affiliated to the network. It has thus the following functions (see Figure 2):

- it maintains the list of servers affiliated to the network, together with a general profile for each of them (containing the languages dealt with by the server, the categories of document it may provide, etc). To this end, each server is described by its name, its address (URL), and a flag describing its current status;
- it should be able to answer any query from each server belonging to the network when they have to know the list of available servers, the information attached to them and their current status;
- it updates the local database of the specialised servers through the network, each time a change has been made to the network description or when something has changed on the part of a given server;
- it is accessible through a simple form-based GUI, to allow the person responsible for the network (the "network master") to add a new server, modify the characteristics of a given server and/or disconnect a server from the network. At this point, no automatic processing will be made from the servers to NMU to ensure full control of the quality and coherence of the information ;

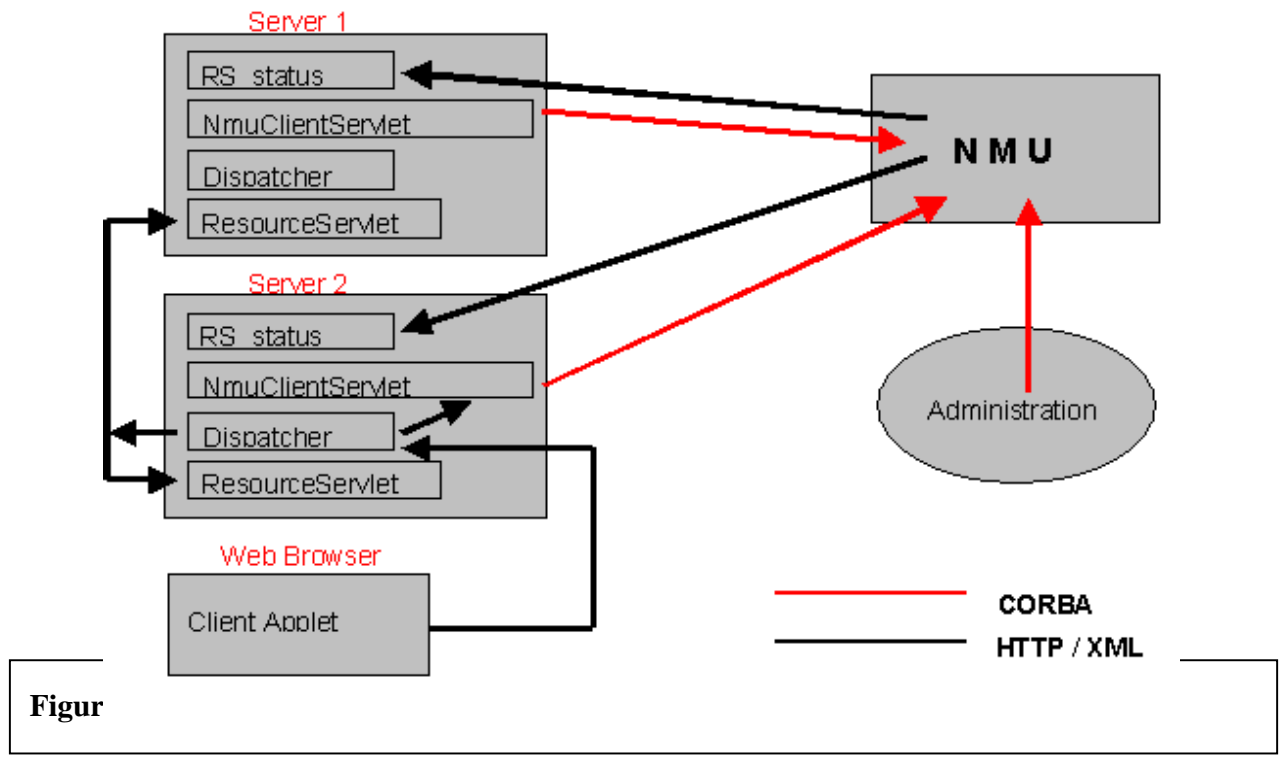

**Figur**

Because of its central position in the architecture, the NMU is subject to several constraints, which are to condition its implementation. In particular, it has to be reliable as to its content since it is in charge of providing each server with the proper information to make it communicate with the others.

There are three related but somehow different problems concerning clients, servers and the NMU:

1. we must allow servers to communicate with clients;
2. these servers have to communicate to each other;
3. we also have to implement a mechanism allowing these servers to communicate with the NMU.

These points will be described in detail in the following sections.

## 4. Information exchange

In order to solve client/server and server/server communication, we have used Java's "servlets" and CORBA.

Servlets are the equivalent of CGI programs, but as they run in a single Java Virtual Machine (JVM), they don't have the per-request overhead of starting a new process per request. Servlets persist across multiple requests and hence can easily maintain states such as open database connections over these requests. In addition, there are all the usual advantages of Java : platform portability, safety (a buggy servlet won't bring down the server), enterprise integration (e.g. JDBC, CORBA, RMI), etc. Of course, servlets are associated to an HTTP server.

Servlets allow us to solve some security problems that we may have when using applets.

For instance, all information concerning users, will be stocked as XML documents. All these XML documents are created at the user's level by a Java applet. Because of security reasons, Java applets may not create files directly on the server. As a consequence, the XML document cannot be written on the server.

Thus, the client's applet and the server communicate via a servlet. We should note that, the client's applet and the servlet allowing communication with the server, must be "physically" downloaded from the same URL. Obviously, the protocol used by the servlet in the framework of applet/server communication is HTTP.

This communication is made by means of a request, in the very same way as a CGI script. That is, in order for the applet to retrieve the XML document concerning a user – this document contains the user's login and password -, we only have to make the request by using the servlet's related URL. Users database, which are in part administered via the user's interface, are composed by a set of XML documents. All these XML documents may be manipulated simply as files or as a real XML database.

### 4.1. Technical Issues

### 4.1.1. The XML level

All information flowing through the network will be encoded in XML. The encoding system will obviously be defined by DTDs. These DTDs allows to encode user's working spaces (WS), requests on data and meta-data (QL), results for these requests(RS), information related to users(UI) , and so on.

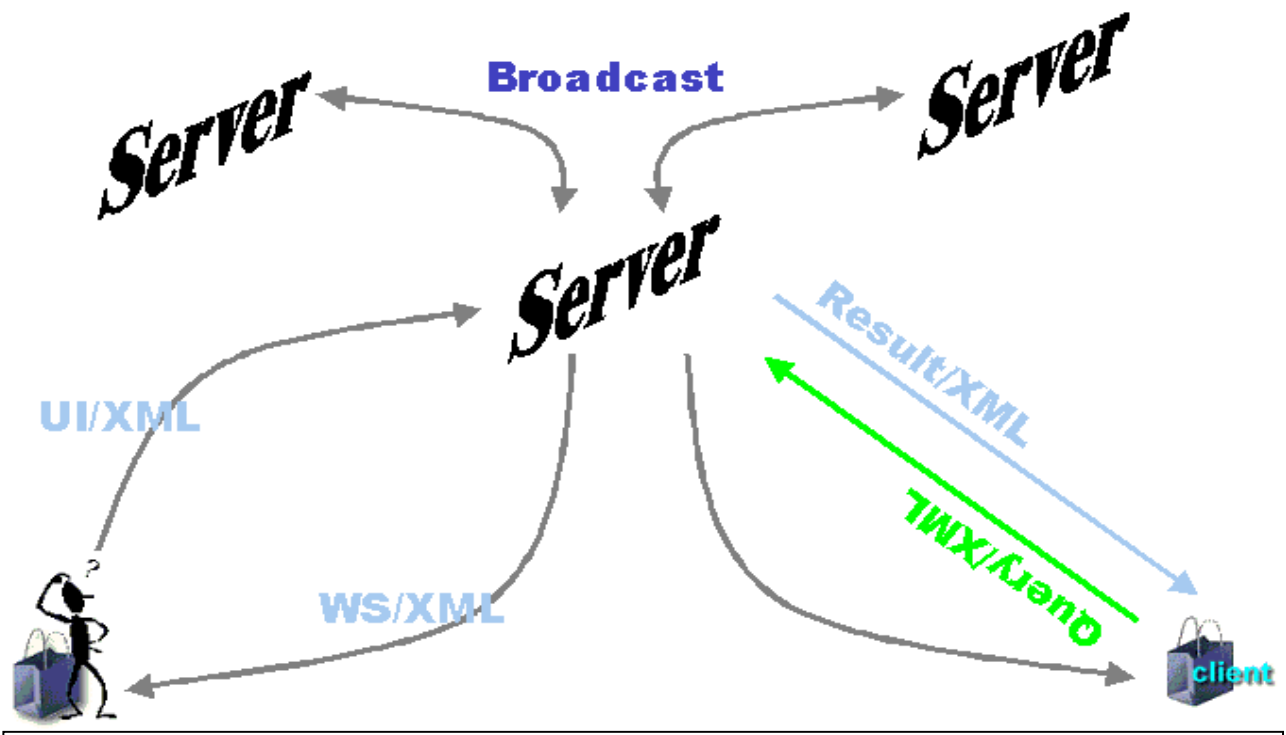

Figure 3. "Information Flow in the Network".

### 4.1.2. The MIME level

Because of efficiency, in particular in the framework of broadcasting requests from a server to another, not all needed information is presented in XML documents. In general on the Internet, data like sound, images, etc,. are encapsulated on a MIME layer. MIME allows to assign a data type (i.e., content-type), as well as to encapsulate heterogeneous data on a single data flow called `MIME/Multipart`: on a single channel, one single connection is enough in order to exchange multiple data.

### 4.1.3. The HTTP level

We use the HTTP protocol as the main communication protocol. Using HTTP is extremely simple (two single Java instructions allow to open an HTTP connection). However, http being a "stateless" protocol., each time a new page is loaded, the user is effectively disconnected from the server and it keeps no information allowing to know who was the user and what he was doing.

The global scenario is:
1. Open the client/server connection;
2. send a request from the client to the server;
3. send a response from the server to the client;
4. close the client/server connection.

Thus, even after logging into a site, each page accessed must pass the username and the password back to the server to verify the user's right to access the page. The client application (the browser) and the server application (the Web server) have no concept of local variables, local method calls, or objects. This problem is solved in part by using `MIME/Multipart`.

### 4.2. Query broadcast and result processing

On the server's side, all requests related to meta-data are analysed by means of a "local transaction". From a technical point of view, a Transaction is integrated within a Java's Servlet Session.

### 4.2.1. Transaction

Every access to the resources database is made by means of a local transaction instance. In the framework of a transaction, all accesses to the resources database are considered to be persistent until the end of this Transaction.

A Transaction may be open or closed:
- `open()`, while receiving a request, for example, or during a database update operation;
- `close()`, in general by a user's request or because a "timeout" operation.

We can also test any time, if the transaction is active by using the `isopen()` operation.

Managing an XML database is also done by means of a transaction. This transaction allows to control all current operations:
- `commit()`, in order to execute an operation
- `abort()`, to stop an operation.

### 4.2.2. Transaction/Query Servlet

Activating a Transaction by using a Java's Servlet allows to simulate the persistence of the client / server connection via a `ServletSession` provided by the JSDK[1] API. This persistence allows a user to progressively keep all results he is interested in. Thus, even if the evaluation of a request returns a high number of results, the user doesn't need to have a high storage or process capacity, in order to "navigate" inside the result's set.

## 4.3. Security aspects

Two main security threats can be identified in the proposed architecture:
- Controlling access to the NMU and especially to the administration methods.
- Securing the communications between servers.

This basically means that we need to:
- Identify trusted clients on the NMU;
- Identify users on servers so only the authorised ones get access to protected resources;
- Make sure that no user can connect as someone else, in order to get specific privileges;
- Make sure that intercepting the data exchanged between servers is useless.

Identification of users to the servers is implemented through a classic login and password procedure. Our purpose here is to present different methods that should be implemented and tested in order to secure the transactions taking place in the Network, on the one hand, between servers themselves, and on the other one hand, between servers and the NMU.

The methods presented here are all based on well known technologies largely used on the Internet, and available for systems implemented in the Java language.

### 4.3.1. Securing access to the NMU (Network Management Unit)

Some of the methods provided by the NMU IDL interface are used for administrating the NMU database, and so have to be protected in a way a fake client implemented by some "entrusted" people could not execute them, and thus trash the NMU database.

Therefore, we have to implement a system allowing only trusted clients to call these methods.

Two options are available for securing access to the NMU. The first one, the CORBA Security service, is implemented as a standard CORBA service. The second one is based on a password system.

## 5. Implementing the proposed architecture.

---

[1] Java Servlet Development Kit, integrated within JDK 1.2

## 5.1. The SILFIDE Project[1].

SILFIDE is a tool for sharing, congenially and thoughtfully, a knowledge on different aspects of the French language. It consists in a network of data processing servers together with the necessary support.

The aim of SILFIDE is not to integrate the totality of the contents (corpora, glossaries, tools) of the available resources within an academic community, but to allow any researcher to be informed of the existence of such contents, to get a relatively precise idea of them and to be informed of the methods of access. In the case of resources which are widely used or which do not raise particular problems when accessed, SILFIDE will be able to propose the automatic transfer of the corresponding data.

Thus, SILFIDE is a help provided to all laboratories of the French-speaking community and to those who are interested in the study or the automatic treatment of French language. In this respect, French has to be the main language of our project. On the one hand, most of the data available on SILFIDE will be in French or associated to equivalent data in French (in the case of a parallel corpus for instance). On the other hand, French will be the meta-language associated to the management of the resources either at the level of their documentation or at the level of the access interface to the corpus. However, a description of the server in other languages (English or German for instance) would be useful.

In the beginning, SILFIDE should be able to answer the following questions which may be raised by an user:

- What are the available data ?
- Where are they available and under which format ?
- What are the conditions of access ?
- How (and by whom) have these data been compiled ?
- What is the validation degree of the Resources ?
- What are the tools available to manipulate these resources ?

Besides the function of access to linguistic resources, it might be interesting to propose directly accessible "on-line" tools to users who do not have access to an elaborate computer environment. Concordances can then be thought out for a set of selected texts together with elementary lexical statistics (frequencies, reduced deviations, etc.)

Moreover, SILFIDE should be helpful by compiling (and possibly by documenting) the tools available in the field of textual resources manipulation. It can be a matter of encoding data but also libraries functions dedicated to normalised data. These different additional functions will have to be progressively integrated to the successive versions of the SILFIDE server.

## 5.2. The MLIS/ELAN Project[2].

Companies, research teams and individuals involved in language engineering or in undertakings such as translating, dictionary making or philology require LARGE corpora, lexica and similar electronic resources. On the other hand such linguistic databases have already been created for most European languages. It can be observed that the need exists, the product exists and yet the two have until now rarely been brought together.

This paradox can largely be explained by the fact that the language resources in question often cannot be accessed EASILY and usually cannot be exploited using STANDARDISED procedures.

Most existing European databases are not widely publicised, can only be accessed on site and are designed only for the immediate needs of the owner or producer. As a result, the software is often neither user-friendly nor portable, with documentation and support being non-existent or scant. In addition, the lack of a clear pricing means that each potential user must negotiate individually with the provider.

Standardised procedures postulate that similar data is made available in a similar format and that a common query system can operate on it. Appreciable progress was made during the 90's regarding the standardisation of formats through projects such as the TEI. Concerning the second requirement nothing has been achieved so far.

Two non-profit international associations — PAROLE and TELRI — have joined forces to undertake the ELAN project in order to link the existing resources with their potential users throughout Europe.

In order to serve the electronic multilingual resource market ELAN plans:

- to reinforce or, where necessary, create international standards by designing a common query language (ELAN-CQL) and by providing standardised resources for the following languages : Albanian, Belgian French, Belorussian, Bulgarian, Catalan, Croatian, Czech, Danish, Dutch, English, Estonian, Finnish, French, German, Greek, Hungarian, Irish, Italian, Latvian, Lithuanian, Polish, Portuguese, Romanian, Russian, Serbian, Slovakian, Slovene, Spanish, Swedish and Uzbek.

- to operate a user community network with active awareness-raising measures, a clear copyright policy, user support, e-mail user groups, etc.

One of the objectives of the MLIS/ELAN project is to define a proper software environment through which it might be possible to access and/or distribute linguistic resources, which would be spread among different servers.
There are different reasons why to go about in this fashion.
First, it should be considered that there already exist several sites around the world (most of the European ones being represented in ELAN) which actually act as brokers for different types of linguistic resources. The idea emerging behind the sole period of the ELAN project is actually to be able to bring together the corresponding resources, in order to provide any user with a global access to these. In some cases this will lead to more coherence between these databases since redundancies might be detected or complementary resources (e.g. parallel texts) put into correspondence.
Second, the classical view of a centralised database containing all the information in a given domain is far from applicable to linguistic resources where, because of their intrinsic diversity (prose, theatre, poetry, newspaper articles, dictionaries, historical documents etc.), there is a need for them to be created and above all maintained at a place where there is the competence to do so. In particular, the encoding background adopted within ELAN, that is SGML (and its subset XML), allows one to continuously enrich documents with specific linguistic annotations like part of speech (POS) tags or proper names (to quote only two possibilities).
Finally, there can be specific constraints that can preclude some given resources to be deported to another site than the site which has originally created them. In particular, some of the partners within ELAN have specific agreements with publishers, which express strong conditions on the actual distribution of electronic files. It is thus more sensible not to take the risk of hampering the agreement by overly spreading the corresponding contents. In the case of the ELAN network, each resource is only accessible through specific queries which can thus be controlled as to their actual applicability.

## 5.3. Information Exchange.

As indicated before, all information flowing throw the network is encoded in XML. In SILFIDE we have defined a "SILFIDE Interface Language" which is obviously specified by a DTD. This DTD is composed by a set of four modules defining working spaces (WS), requests on meta-data (QL), results for these requests (RS), and information related to users (UI). The MLIS/ELAN project use the same DTD but an extension mechanism adds a new DTD called "Elan Query" that specifies the ELAN "Common Query Language" Language.
The MLIS/ELAN network, based on the SILFIDE architecture, provides a simple but general way to interface any native database management system.
We just have to specify an interface "ElanQueryHandler" for handling :
1.  Connection and Transaction;
2.  Query;
3.  Result Set and Enumeration.
To "plug" the database into an ELAN server it is necessary to provide a driver that implements the "ElanQueryHandler" interface functions.
To optimise the transaction efficiency of the results and to avoid an overload of the network, we have implemented a "double cache" mechanism to store and send the result sets of the remote and local servers. The user has the possibility to specify the maximum size of the result set he can handle at a time. The size of the cache is set by the administrator of the server.
Figure 4 shows the information exchange process in the MLIS/ELAN network. We have the following steps :
1.  A user send a CQL query to his local server;
2.  The broadcast servlet intercepts and broadcast the query to each selected remote server (and even to the local server);
3.  Each server establish a connection to the database, through the ElanQuery driver, and manage the transaction of the result with the server ("remote cache");

4. Each remote server send the results, set by set, to the local server ("local cache") and the local server "distils" the results to the client.

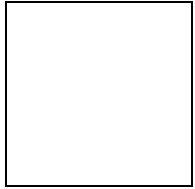

## 6. Concluding Remarks.

## 7. Bibliography.

**Figure 4. "Information Exchange in the MLIS/ELAN Project".**

## 8. Annex "A" The "SILFIDE Interface Language" DTD.

```
<!DOCTYPE sil [

<!--
        Silfide Interface Language DTD
        Version 0.5
        Draft: Thu Jan  7 22:22:57 MET 1999
        Author: Patrice Bonhomme
-->

<!-- Typical usage:
        <!DOCTYPE sil SYSTEM "http://www.loria.fr/projets/XSilfide/dtd/sil.dtd">
        <sil>
         ...
        </sil>
-->

<!-- SIL Module declaration -->

<!ENTITY % SIL.ws SYSTEM "ws.dtd">
<!-- PUBLIC "-//Silfide//DTD SIL Work Space 0.5 Draft 19990106//EN" -->
<!-- SYSTEM "http://www.loria.fr/projets/XSilfide/dtd/ws.dtd" -->
%SIL.ws;

<!ENTITY % SIL.ui SYSTEM "ui.dtd">
<!-- PUBLIC "-//Silfide//DTD SIL User Information 0.5 Draft 19990106//EN" -->
<!-- SYSTEM "http://www.loria.fr/projets/XSilfide/dtd/ui.dtd" -->
%SIL.ui;

<!ENTITY % SIL.ql SYSTEM "ql.dtd">
<!-- PUBLIC "-//Silfide//DTD SIL Query Language 0.5 Draft 19990106//EN" -->
<!-- SYSTEM "http://www.loria.fr/projets/XSilfide/dtd/ql.dtd" -->
%SIL.ql;

<!ENTITY % SIL.rs SYSTEM "rs.dtd">
<!-- PUBLIC "-//Silfide//DTD SIL Result Set 0.5 Draft 19990106//EN" -->
<!-- SYSTEM "http://www.loria.fr/projets/XSilfide/dtd/rs.dtd" -->
%SIL.rs;

<!-- SIL extensions default declaration -->

<!ENTITY % SIL.x.dtd "" >
%SIL.x.dtd;

<!ENTITY % SIL.x.ent "" >
%SIL.x.ent;

<!-- SIL extension element names -->
<!ENTITY % SIL.x.e ''>

<!-- SIL extension attribute names -->
<!ENTITY % SIL.x.a ''>

<!-- SIL module element names -->
<!ENTITY % SIL.module.e '%SIL.x.e; ws | ui | ql | rs'>

<!-- SIL module attribute names -->
<!ENTITY % SIL.module.a ''>
```

```
<!-- SIL Document structure -->

<!ELEMENT sil (uid, (%SIL.module.e;)+)>
<!ATTLIST sil type    (%SIL.x.a; workspace | user | query | resultset)
"workspace"
               crdate  CDATA #IMPLIED
               update  CDATA #IMPLIED
               lang    CDATA #IMPLIED
               sid     CDATA #REQUIRED
               version CDATA #FIXED "0.5">

<!-- lang : [RFC1766] language value
     sid  : Silfide server identification -->

<!-- =================== -->
<!-- User IDentification -->
<!-- =================== -->

<!-- validated by the Silfide server admin. -->

<!ELEMENT uid   (login, passwd, access)>
<!ATTLIST uid    type (user | provider | both) "user">

<!ELEMENT login    EMPTY>
<!ATTLIST login    id ID #REQUIRED>

<!ELEMENT passwd   (#PCDATA)>

<!ELEMENT access   (default,group*)>

<!ELEMENT default   EMPTY>
<!ATTLIST default   group IDREF #IMPLIED>
<!-- cross reference to a <group id="..."> element -->

<!ELEMENT group    EMPTY>
<!ATTLIST group    id ID #REQUIRED>
<!-- id : group identification key : inalf, lingua, etc. -->

]>
```